\documentclass[runningheads]{llncs}

\usepackage{graphicx}
\usepackage{amsmath}
\usepackage{amssymb}
\usepackage{booktabs}
\usepackage{xcolor}
\usepackage{multirow}
\usepackage{todonotes}
\usepackage{tabularx}
\usepackage{url}
\usepackage{comment}
\usepackage{adjustbox}
\usepackage{color}

\definecolor{CP}{rgb}{0.,0.,1}

\newcommand{\printfnsymbol}[1][\value{footnote}]{\footnotemark[#1]}

\newcommand\asteriskfill{\leavevmode\xleaders\hbox{$\ast\ $}\hfill\kern0pt}

\usepackage{authblk}

\begin{document}

\title{Longitudinal Quantitative Assessment of \\ COVID-19 Infection Progression \\ from Chest CTs}

\titlerunning{Longitudinal Quantitative Assessment of COVID-19}

\author{Seong Tae Kim\inst{2,}\thanks{\textit{First two authors contributed equally to this work.}}, Leili Goli\inst{3,}\printfnsymbol, Magdalini Paschali\inst{1}, Ashkan Khakzar\inst{1}, Matthias Keicher\inst{1}, Tobias Czempiel\inst{1}, Egon Burian\inst{4}, Rickmer Braren\inst{4}, Nassir Navab\inst{1,5}, Thomas Wendler\inst{1}}
%index{Kim, Seong Tae}
%index{Goli, Leili}
%index{Paschali, Magdalini}
%index{Khakzar, Ashkan}
%index{Keicher, Matthias}
%index{Czempiel, Tobias}
%index{Burian, Egon}
%index{Braren, Rickmer}
%index{Navab, Nassir}
%index{Wendler, Thomas}

\institute{
Computer Aided Medical Procedures, Technical University of Munich, Germany
\and
Department of Computer Science and Engineering, Kyung Hee University, Korea
\and
Department of Computer Engineering, Sharif University of Technology, Iran
\and
Department of Diagnostic and Interventional Radiology, Technical University of Munich, Germany 
\and
Computer Aided Medical Procedures, Johns Hopkins University, USA
}
\authorrunning{Kim and Goli et al.}

\maketitle            
\begin{abstract}
Chest computed tomography (CT) has played an essential diagnostic role in assessing patients with COVID-19 by showing disease-specific image features such as ground-glass opacity and consolidation. Image segmentation methods have proven to help quantify the disease and even help predict the outcome. The availability of longitudinal CT series may also result in an efficient and effective method to reliably assess the progression of COVID-19, monitor the healing process and the response to different therapeutic strategies. In this paper, we propose a new framework to identify infection at a voxel level (identification of healthy lung, consolidation, and ground-glass opacity) and visualize the progression of COVID-19 using sequential low-dose non-contrast CT scans. In particular, we devise a longitudinal segmentation network that utilizes the reference scan information to improve the performance of disease identification. Experimental results on a clinical longitudinal dataset collected in our institution show the effectiveness of the proposed method compared to the static deep neural networks for disease quantification.
\keywords{Longitudinal Analysis \and COVID-19 \and Disease Progression}
\end{abstract}

\section{Introduction}
The Coronavirus Disease 2019 (COVID-19) has infected more than 113 million people worldwide (as of February 28th, 2021\footnote[1]{https://coronavirus.jhu.edu/map.html}) and caused more than 2.52 million deaths. Although many cases present only mild symptoms, some of them evolve into serious illnesses that require intensive medical treatment or lead to death~\cite{harmon_artificial_2020}. 

Chest computed tomography (CT) has played an essential diagnostic role in the assessment of patients with COVID-19 by showing specific image patterns such as ground-glass opacity (GGO), crazy paving, and consolidation~\cite{wong2020frequency}.
Several studies have proposed to automatically analyze COVID-19 infection on chest images with deep learning~\cite{fan2020inf,zhou2020rapid,wang2020noise,khakzar2021towards,khakzar2021explaining}. However, it is still challenging to automatically identify and quantify image findings associated with COVID-19 due to the subtle anatomical boundaries, pleural-based location, and variations in size, density, location, and texture~\cite{lei2020ct,shi2020radiological,zhou2020ct}. Moreover, it is vital to develop an effective method to reliably assess the progression of COVID-19 and response to therapy \cite{feng_dynamic_2020,wu_clinical_2020}. Given the availability of multiple therapy options and to understand anatomical changes during the healing process, a longitudinal evaluation of CT images can be beneficial~\cite{feng_dynamic_2020,huang_dynamic_2021}.

Only a few studies have devoted to developing a deep learning-based approach to assess the progression and response to therapy from longitudinal CT scans~\cite{zhang2020dabc,pu2021automated}. Zhang et al. propose a static segmentation model to classify the patients into mild and severe patients based on the features extracted from longitudinal CT scans~\cite{zhang2020dabc}. Pu et al. propose a framework consisting on the following steps (1) segment the lung boundary and vessels, (2) register the boundary between serial scans using deformable registration, (3) identify regions with morphological changes due to the disease, and (4) assess disease progression~\cite{pu2021automated}. The registered longitudinal CT scans are used to generate a heatmap visualizing the difference in diseased vs. healthy areas between scans. Yet,  the identification of affected regions is performed in a static way, i.e., the information between serial scans is not taken into account. Besides, they do not differentiate image features of consolidation and GGO, even though these pathologies provide different information of infection in COVID-19 cases~\cite{zhou_ct_2021}.

In this paper, we propose a novel framework to identify infection at a fine-grained level (identification of healthy lung, consolidation,  GGO, and pleural effusion) by leveraging spatio-temporal cues between longitudinal scans and visualize the progression of COVID-19. In particular, we devise a longitudinal segmentation network that utilizes the reference scan information to improve the performance of disease segmentation. Even though longitudinal scans share structural information, differences exist due to the progression of the disease. We investigate and propose ways to use this information during the segmentation based on a dataset collected during the first COVID-19 wave of our institution. The following is the summary of the main contributions.
\begin{itemize}
\item To the best of our knowledge, this is the first study to explore the longitudinal segmentation of CTs of COVID-19 patients. By designing a deep network to use the information provided in the reference scan, we show that the performance of segmentation can be improved compared to the static segmentation model. Our method shows promising results with limited data.
\item We propose a framework to analyze the progression of COVID-19 infection over time, which is crucial for the course of the disease and the patient’s recovery.
\item We present comprehensive analysis and ablation studies to verify our longitudinal analysis framework’s design choice.
\end{itemize}

\section{Methodology}
\begin{figure}[t]
\centering
  \includegraphics[width=0.95\textwidth]{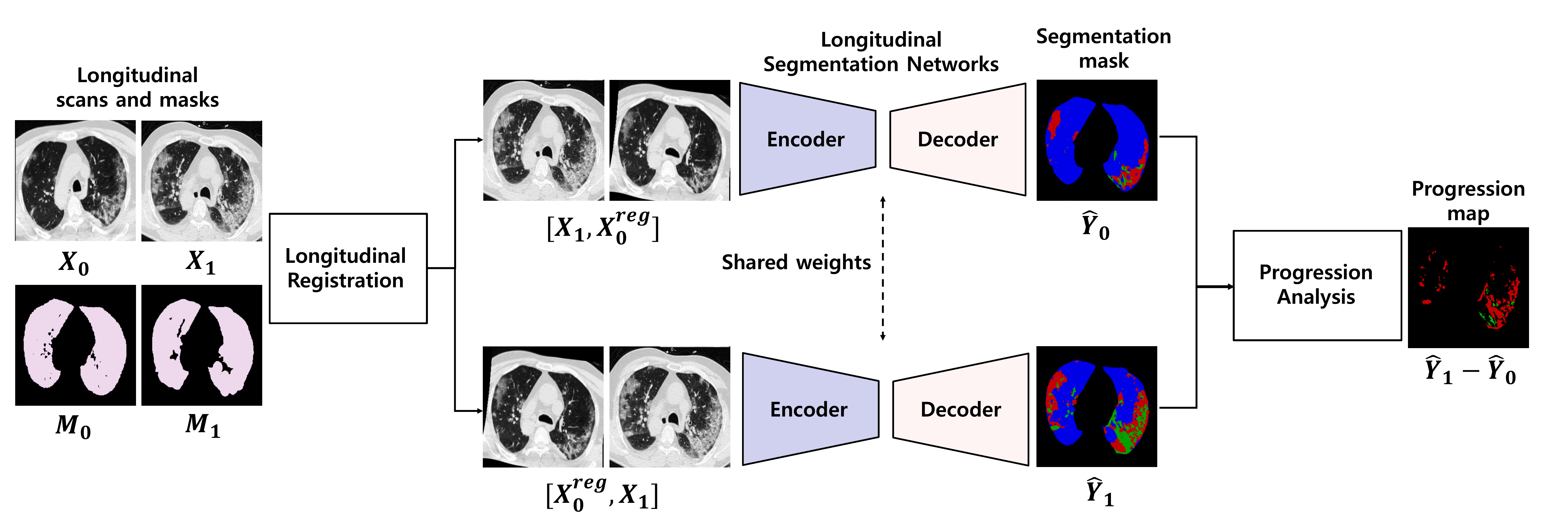}
  \caption{Progression analysis framework: The framework is comprised of three modules: Longitudinal registration, longitudinal fine-grained segmentation to identify pathologies, and progression analysis. The inputs are two consecutive CT scans (t=0: reference scan and t=1: follow-up scan).}
  \label{fig:methods}
\end{figure}

This section describes our approaches for incorporating spatio-temporal features into the framework of segmentation and progression quantification. Figure~\ref{fig:methods} shows overall framework which consists of the following three components: 1) longitudinal registration, 2) longitudinal segmentation, and 3) progression analysis. The analyses are performed based on two different time point scans.

\subsection{Longitudinal Registration}
Due to the nature of chest CT scans, the initial volumes of data between different time points are highly misaligned. Aspects like patient positioning, variations of the imaging parameters or devices, different phases in the breathing cycle, and the disease progression are the main reasons. This misalignment cannot be described as a linear transformation composed of translation and rotation between the time points of data and can only be expressed through non-linear transformations. The misalignment can make the network incapable of using the longitudinal information present between different time points of data. 

As a solution for this problem, we utilize a deformable registration algorithm~\cite{lowekamp2013design} where a BSpline Transform is defined using a sparse set of grid points overlaid onto the fixed domain of the image domain to deform it. Using this algorithm we register the reference scan lung mask $M_0$ to the follow up scan lung mask $M_1$ and this transform function is defined by $R_{M_0 \rightarrow M_1}(\cdot)$. Based on this function, the transformations are applied to the respective CT-scans as ${X}^{reg}_{0}=R_{M_0 \rightarrow M_1}(X_0)$. Using the lung masks, we avoid registration errors due to the pathological changes in the lung parenchyma while compensating for positioning, breathing phase, and acquisition-related differences.

\subsection{Longitudinal Segmentation}
Due to the challenge of training a 3D model with a limited number of training data, 2.5D approaches~\cite{aslani2019multi,zhang2019multiple,roy2019quicknat,denner2020spatio} have shown state-of-the-art results on various medical segmentation problems. For the 3D approach, it is challenging to directly process a full 3D volume by using current GPU memory limitations~\cite{roy2019quicknat}, which forces people to operate on 3D patches~\cite{wachinger2018deepnat,hashemi2018asymmetric}. However, patch-wise training limits the overall spatial context for accurate semantic segmentation. 

In this study, we adopt the 2.5D approach of~\cite{denner2020spatio} with a fully convolutional (FC) DenseNet~\cite{jegou2017one} as a baseline 2D segmentation model. The FC DenseNet consists of a downsampling path (Encoder) with 5 Transitions Down blocks, each with 4 layers and an upsampling path (Decoder) with 5 Transitions Up blocks, each with 4 layers. The model is trained for all three views (coronal, sagittal, and axial view). At test time, for each given voxel, the segmentation is conducted on all three orthogonal views. Afterward, the predicted probability of a given voxel is averaged among views to assign a final predicted probability. 

We extend the aforementioned 2.5D segmentation to deal with longitudinal information by modifying its architecture. To capture subtle spatio-temporal cues, we concatenate two registered scans from two different time-points as input for the longitudinal segmentation network: $[X^{reg}_0,X_1]$ for segmenting pathologies on $X_1$ or $[X_1, X^{reg}_0]$ for $X_0$. This enables the segmentation network to capture temporal changes as shown in Figure~\ref{fig:methods}. For subjects who have more than two scans, we select consecutive scans for the segmentation in each step.

\subsection{Progression Analysis}
To monitor the progression of the COVID-19 infection and the response to therapy, we extract the segmented pathologies from consecutive longitudinal CT scans and quantify the volume differences between them.
Our approach is capable of quantifying all combinations of changes between different pathologies and healthy lung parenchyma.
In this work’s scope, we define two classes as consolidation and non-consolidation since consolidation has shown to be a robust biomarker for COVID-19~\cite{li2020clinical}.
The progression of consolidation is computed by subtracting two registered longitudinal CT-scans. The resulting residual voxel values could be either positive or negative. Positive voxels indicate that a healthy, GGO or pleural effusion region progresses to consolidation (Progression), and negative voxels suggest that an area recovers from the severe infection of consolidation (Recovery). 

\subsection{Training with Progression Information}
For the optimization of our model, we define an overall loss function combining a segmentation loss $\mathcal{L}_{seg}$ and a progression loss $\mathcal{L}_{prog}$. Note that as shown in Fig.~\ref{fig:methods}, the outputs of our model consist of the segmentation masks for reference (t=0) and follow-up (t=1) scans and the subtracted volume between reference and follow-up scans for presenting progression of the COVID-19 infection. The overall loss is defined as:
$\mathcal{L} = \mathcal{L}_{seg} + \mathcal{L}_{prog}.$

The segmentation loss is defined as $\mathcal{L}_{seg}=L_{MSE}(Y_0,\hat{Y_0})+L_{MSE}(Y_1,\hat{Y_1})$ where $L_{MSE}$ denotes a mean squared error loss \cite{zhang2019multiple,denner2020spatio}. In our evaluation, this metric yielded better results in comparison to the dice score. $Y_0$ and $\hat{Y_0}$ denote a ground truth pathology segmentation map and a predicted segmentation mask for t=0 scan, respectively. $Y_1$ and $\hat{Y_1}$ denote a ground truth pathology segmentation map and a predicted segmentation mask for t=1 scan.

The progression loss is defined as $\mathcal{L}_{prog}=\mathcal{L}_{MSE}(Y^{con}_1-Y^{con}_0,\hat{Y}^{con}_1-\hat{Y}^{con}_0)$ where $Y^{con}$ denotes a ground truth consolidation map and $\hat{Y}^{con}$ denotes a predicted consolidation map. In those binary maps, consolidation is mapped to 1 and non-consolidation to 0. As explained above, the progression map is calculated by $Y^{con}_1-Y^{con}_0$. Note that, the progression loss does not minimize the distance between the two segmentations. Instead it explicitly uses the structural changes of pathologies through times as cues for modeling the optimization. In other words, if there are large changes over time, the progression loss encourages the model to predict those large changes also in the segmentation.

\section{Experiment Setup}
\subsection{Dataset}
To our knowledge there is no publicly available longitudinal CT dataset for COVID-19. Accordingly, to evaluate the proposed method, we used an in-house clinical dataset which consists of longitudinal low-dose CT-scans from 38 patients (64$\pm$18 years old, 16 females, 22 males) with positive PCR from the first COVID-19 wave (March-June 2020). 28 patients had two scans and 10 had three. The CTs were separated 17$\pm$10 days (1-43 days) and were taken at admission and during the hospital stay (33$\pm$21 days, 0-71 days). 8 patients of the 38 died; 30 recovered from COVID-19, 20 of them needing intensive care. All scans were performed in-house using two different CT devices (IQon Spectral CT and iCT 256, Philips, Hamburg, Germany) with the same parameters (X-ray current 140-210 mA, voltage 120 kV peak, slice thickness 0.9mm, no contrast media) and covered the complete lung. The data was collected retrospectively with the approval of the institutional review board of our institution (ethics approval 111/20 S-KH).

The dataset was annotated at a voxel-level by a single expert radiologist (5 years experience), generating lung masks (lung parenchyma vs. other tissues) and pathology masks including four classes: healthy lung (HL), GGO, consolidation (CONS), and pleural effusion (PLEFF). For segmentation, the radiologist used the software ImFusion Labels (ImFusion, Munich, Germany).

The dataset was split into a training set of 16 patients (37 volumes) and an independent test set of 22 patients (49 volumes). From the training set, 12 patient scans are used for model training and 4 patient scans are used for validation. The model is finally evaluated on the unseen test set. 

\subsection{Implementation Details}
The raw CT volumes highly vary in intensity range, size, and alignment. Therefore, we perform the following pre-processing steps on the raw volumes to enable effective use of the longitudinal data:

\noindent \textbf{Cropping.}
Since different body regions can be presented between time points and patients, we crop the volumes to the lung regions, using the manually-annotated lung masks.

\noindent \textbf{Clipping and Normalization.} To alleviate different intensity ranges among CT-scans, intensity values outside the range $(-1024, 600)$ are clipped and then min-max normalization is performed on each volume. 

\noindent \textbf{Resizing.} After Cropping, resulting volumes vary in size in all three dimensions, ranging from 100 pixels to 580 pixels. Therefore all volumes are resized to a fixed size of 300$\times$300$\times$300 with 300 being the median among the cropped-volume sizes. 

\noindent \textbf{Slicing and Removing Empty Slices.} Finally, the volumes are sliced in each of the three dimensions to 300 slices, generating sagittal, coronal, and axial views of the lung. Slices that have a voxel-value variation smaller than $0.001\%$ between their maximum and the minimum value are considered empty and are removed. 

\noindent \textbf{Model Training.} For training, Adam optimizer~\cite{kingma2014adam} with a learning rate of 0.0001 and a decay rate of 0.1 for every 50 steps was used. The model was trained over 100 epochs with early stopping if no decrease in the validation loss was computed for 5 epochs. 
Our method was implemented in PyTorch 1.4 and our models were trained on an NVIDIA Titan V 12GB GPU using Polyaxon\footnote[2]{https://polyaxon.com/}. The source code is publicly available\footnote[3]{https://github.com/lilygoli/longitudinalCOVID}. Our longitudinal model had 1.3752M parameters in comparison to its static counterpart 1.3748M.

\section{Results and Discussion}
\begin{table}[t]
	\centering
	\caption{Comparison of different methodologies for segmenting CoViD-19 infection on the independent test set. Dice similarity coefficient is used for metric. The average and standard error are calculated. $^*$ denotes the case that the difference with the proposed method is statistically significant (\textit{p}$<$0.05). HL, CONS, GGO, PLEFF denote healthy lung, consolidation, ground-glass opacity, and pleural effusion, respsectively.}
    {
    \resizebox{\linewidth}{!}{
    \begin{tabular}{c|c|c|c|c}
        \toprule
        {\textbf{Method}} & \textbf{HL} & \textbf{CONS} & \textbf{GGO}& \textbf{PLEFF}\\
        \hline
        \textbf{Static Network} & 0.796$\pm$0.021$^*$ & 0.322$\pm$0.031$^*$  & 0.380$\pm$0.028$^*$ & 0.210 $\pm$ 0.033$^*$\\ 
        
        \textbf{Longitudinal Network} &  \multirow{2}{*}{0.835$\pm$0.019} &  \multirow{2}{*}{0.402$\pm$0.034} &  \multirow{2}{*}{0.435$\pm$0.029$^*$} &  \multirow{2}{*}{\textbf{0.266$\pm$0.041$^*$}}\\ 
        \textbf{(without progression loss)} & & & & \\
        
        \textbf{Proposed} & \textbf{0.837$\pm$0.022} & \textbf{0.406$\pm$0.035} & \textbf{0.447$\pm$0.030} & 0.246$\pm$0.040\\ 
        \bottomrule
    \end{tabular}}
    } 
    \label{Table_Comparison_Main}
\end{table}

\subsubsection{Effectiveness of Longitudinal Segmentation}
First, we conduct comparative experiments to verify the effectiveness of our longitudinal segmentation method. In Table~\ref{Table_Comparison_Main}, we compare our method with a static network~\cite{zhang2019multiple} based on FC-DenseNet~\cite{jegou2017one} and with a longitudinal network without progression loss. This simpler longitudinal network has the same architecture as our proposed one, i.e., it concatenates longitudinal CT scans as an input for the segmentation model, but it is trained using only the segmentation loss.
As shown in Table~\ref{Table_Comparison_Main}, both longitudinal networks achieved a higher Dice Similarity Coefficient (DSC) than the static network. The difference was statistically significant for all classes (\textit{p}$<$0.05 by paired t-test~\cite{altman1990practical}). This implies that using longitudinal information from the reference CT scan is informative to segment pathology on the target CT scan. In our longitudinal network with progression loss, the DSC was further improved for HL, CONS, and GGO. But for PLEFF, the performance slightly decreased. This can be attributed to the fact that the progression loss encourages the model to focus on CONS rather than on PLEFF. Additionally, PLEFF is a challenging, under-represented class in our dataset (only 2.17\% voxels). 

\subsubsection{Effect of Longitudinal Registration.}
\begin{table}[t]
	\centering
	\caption{Ablation study to investigate the effectiveness of the registration and using temporal information in our longitudinal segmentation model. Dice similarity coefficient is measured on the independent test set.}
    {
    \resizebox{\linewidth}{!}{
    \begin{tabular}{c|c|c|c|c|c|c}
        \toprule
        \textbf{Method} & \textbf{Registration}& \textbf{Long. Input}&\textbf{HL} & \textbf{CONS} & \textbf{GGO}& \textbf{PLEFF}\\
        \hline
        \textbf{Without Registration} & &\checkmark & 0.761 & 0.311 & 0.388 & 0.146\\ 
        \textbf{Static Input}&\checkmark & & 0.774 & 0.327 & 0.224 & 0.160\\
        \textbf{Proposed} & \checkmark & \checkmark& \textbf{0.837} & \textbf{0.406} & \textbf{0.447} & \textbf{0.246}\\

        \bottomrule
    \end{tabular}}
    } 
    \label{Table_Analysis_Registration}
\end{table}
To showcase the importance of registration among the longitudinal scans, we report results with and without deformable registration. As seen in Table~\ref{Table_Analysis_Registration} the performance after registration substantially improves across the board with the increase ranging from 0.07 to 0.10.

\subsubsection{Importance of Temporal Information in Longitudinal Network.}
In this experiment, we highlight the importance of the longitudinal scans for the performance of the model. Specifically, we concatenate two duplicates of the reference scan instead of the reference and follow-up scan as input to our model ('static input'). As shown in Table~\ref{Table_Analysis_Registration}, the longitudinal input (proposed method) outperforms the static input for all classes.

\subsubsection{Progression Analysis.}
Finally, we evaluate our method for progression analysis by comparing with a static network~\cite{pu2021automated}, a longitudinal network with multi-view approach~\cite{birenbaum2017multi} and our model trained without the progression loss.  
As shown in Table~\ref{Table_Progression}, our longitudinal architecture has a 3.4\% increase compared to the static network. The proposed model with the progression loss has a 4.8\% increase with respect to static network. Note that the model using the progression loss significantly outperformed the static network~\cite{pu2021automated} and the longitudinal network with multi-view approach~\cite{birenbaum2017multi} (\textit{p}$<$0.05). Moreover, the progression loss had statistically significant improvement for the recovery and average progression prediction compared to the longitudinal model without the progression loss.  
\begin{table}[t]
	\centering
	\caption{Comparison of different methodologies for predicting progression of CoViD-19 infection on the independent test set. Dice similarity coefficient is used for metric. The average and standard error are calculated. $^*$ denotes the case that the proposed method outperforms the baseline methods with statistical significance (\textit{p}$<$0.05).}
    {
    \begin{tabular}{c|c|c|c}
        \toprule
        \textbf{Method} & \textbf{Recovery} & \textbf{Progression} & \textbf{Average} \\
        \hline
        \textbf{Static Network~\cite{pu2021automated}} & 0.266$\pm$0.030$^*$ & 0.471$\pm$0.021$^*$ & 0.368$\pm$0.015$^*$ \\ 
        
        \textbf{Longitudinal Network} &  \multirow{2}{*}{0.287$\pm$0.031$^*$} & \multirow{2}{*}{0.491$\pm$0.028} &  \multirow{2}{*}{0.389$\pm$0.023$^*$} \\
        \textbf{(multi-view~\cite{birenbaum2017multi})} & & &  \\
        
        \textbf{Longitudinal Network} &  \multirow{2}{*}{0.299$\pm$0.032$^*$} &  \multirow{2}{*}{0.505$\pm0.019$} &  \multirow{2}{*}{0.402$\pm$0.014$^*$} \\ 
        \textbf{(without progression loss)} & & &  \\
        
        \textbf{Proposed} & \textbf{0.327$\pm$0.033} & \textbf{0.506$\pm$0.026}& \textbf{0.416$\pm$0.017}\\ 
        \bottomrule
    \end{tabular}}
    \label{Table_Progression}
\end{table}

\begin{figure}[t]
\centering
  \includegraphics[width=0.98\textwidth]{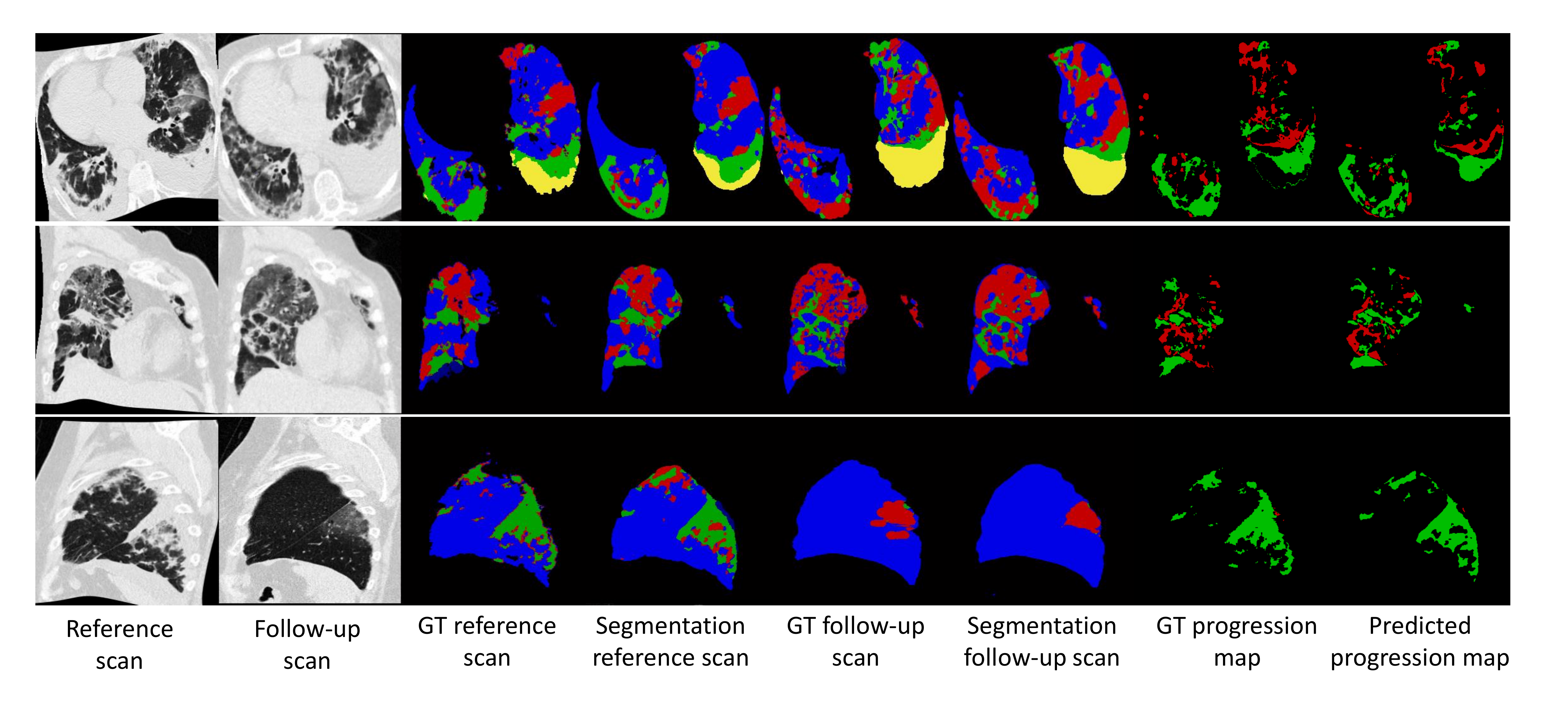}
  \caption{Qualitative results of our method for 3 patients from different views. For the segmentation maps, blue, green, red and yellow denote healthy lung, consolidation, ground-glass opacity, and pleural effusion, respectively. For the progression map, red denotes Progression and the green Recovery.}
  \label{fig:qualitative}
\end{figure}

Figure~\ref{fig:qualitative} showcases qualitative results of the segmentation and progression analysis of our method for all 3 different views. As shown in Figure~\ref{fig:qualitative}, our method successfully provides segmentation and progression maps for both reference and follow-up scans across views and patients. Even the under-represented class of PLEFF is successfully segmented. Regarding the progression, the fine-grained regions of recovered and progressed consolidation are also correctly identified.

\section{Conclusion}
In this work, we proposed a new longitudinal segmentation and progression analysis model for assessing COVID-19 disease over time. Comprehensive experiments were conducted to verify the effectiveness of the longitudinal model. By designing the model to exploit the reference CT scan, our method can achieve higher  progression analysis performance compared to the baseline methods. 

What makes our approach especially interesting is the ability to monitor the development of the infection and healing process in COVID-19, and possibly in other lung diseases. 
It can be used to compute the differences in disease progression of different patient subgroups, such as different COVID-19 variants under the same therapy.
Moreover, it could serve as a quantitative measure to evaluate different therapy approaches. We will further investigate and improve the clinical usability of the method for a larger patient cohort.

\section*{Acknowledgements}
This paper was funded by the Bavarian Research Foundation (BFS) under grant agreement AZ-1429-20C. We would like to thank NVIDIA for the GPU donation.
\label{sec:acknowledgements}

\bibliographystyle{splncs04}
\bibliography{main}

\begin{thebibliography}{10}
\providecommand{\url}[1]{\texttt{#1}}
\providecommand{\urlprefix}{URL }
\providecommand{\doi}[1]{https://doi.org/#1}

\bibitem{altman1990practical}
Altman, D.G.: Practical statistics for medical research. CRC press (1990)

\bibitem{aslani2019multi}
Aslani, S., Dayan, M., Storelli, L., et~al.: Multi-branch convolutional neural
  network for multiple sclerosis lesion segmentation. NeuroImage  \textbf{196},
   1--15 (2019)

\bibitem{birenbaum2017multi}
Birenbaum, A., Greenspan, H.: Multi-view longitudinal cnn for multiple
  sclerosis lesion segmentation. Engineering Applications of Artificial
  Intelligence  \textbf{65},  111--118 (2017)

\bibitem{denner2020spatio}
Denner, S., Khakzar, A., Sajid, M., Saleh, M., Spiclin, Z., Kim, S.T., Navab,
  N.: Spatio-temporal learning from longitudinal data for multiple sclerosis
  lesion segmentation. arXiv preprint arXiv:2004.03675  (2020)

\bibitem{fan2020inf}
Fan, D.P., Zhou, T., Ji, G.P., Zhou, Y., Chen, G., Fu, H., Shen, J., Shao, L.:
  Inf-net: Automatic covid-19 lung infection segmentation from ct images. IEEE
  Transactions on Medical Imaging  \textbf{39}(8),  2626--2637 (2020)

\bibitem{feng_dynamic_2020}
Feng, X., Ding, X., Zhang, F.: Dynamic evolution of lung abnormalities
  evaluated by quantitative {CT} techniques in patients with {COVID}-19
  infection. Epidemiol Infect  \textbf{148}, ~e136 (2020)

\bibitem{harmon_artificial_2020}
Harmon, S.A., Sanford, T.H., Xu, S., Turkbey, E.B., Roth, H., et~al.:
  Artificial intelligence for the detection of {COVID}-19 pneumonia on chest
  {CT} using multinational datasets. Nature Communications  \textbf{11}(1),
  ~4080 (2020)

\bibitem{hashemi2018asymmetric}
Hashemi, S.R., Salehi, S.S.M., Erdogmus, D., Prabhu, S.P., Warfield, S.K.,
  Gholipour, A.: Asymmetric loss functions and deep densely-connected networks
  for highly-imbalanced medical image segmentation: Application to multiple
  sclerosis lesion detection. IEEE Access  \textbf{7},  1721--1735 (2018)

\bibitem{huang_dynamic_2021}
Huang, Y., Li, Z., Guo, H., Han, D., Yuan, F., Xie, Y., et~al.: Dynamic changes
  in chest {CT} findings of patients with coronavirus disease 2019 ({COVID}-19)
  in different disease stages: a multicenter study. Ann Palliat Med
  \textbf{10}(1),  572--583 (2021)

\bibitem{jegou2017one}
J{\'e}gou, S., Drozdzal, M., Vazquez, D., Romero, A., Bengio, Y.: The one
  hundred layers tiramisu: Fully convolutional densenets for semantic
  segmentation. In: CVPR Workshop. pp. 11--19 (2017)

\bibitem{khakzar2021towards}
Khakzar, A., Musatian, S., Buchberger, J., Quiroz, I.V., Pinger, N.,
  Baselizadeh, S., Kim, S.T., Navab, N.: Towards semantic interpretation of
  thoracic disease and covid-19 diagnosis models. arXiv preprint
  arXiv:2104.02481  (2021)

\bibitem{khakzar2021explaining}
Khakzar, A., Zhang, Y., Mansour, W., Cai, Y., Li, Y., Zhang, Y., Kim, S.T.,
  Navab, N.: Explaining covid-19 and thoracic pathology model predictions by
  identifying informative input features. arXiv preprint arXiv:2104.00411
  (2021)

\bibitem{kingma2014adam}
Kingma, D.P., Ba, J.: Adam: A method for stochastic optimization. International
  Conference on Learning Representations (ICLR)  (2014)

\bibitem{lei2020ct}
Lei, J., Li, J., Li, X., Qi, X.: Ct imaging of the 2019 novel coronavirus
  (2019-ncov) pneumonia. Radiology  \textbf{295}(1),  18--18 (2020)

\bibitem{li2020clinical}
Li, K., Wu, J., Wu, F., Guo, D., et~al.: The clinical and chest ct features
  associated with severe and critical covid-19 pneumonia. Investigative
  radiology  (2020)

\bibitem{lowekamp2013design}
Lowekamp, B.C., Chen, D.T., Ib{\'a}{\~n}ez, L., Blezek, D.: The design of
  simpleitk. Frontiers in neuroinformatics  \textbf{7}, ~45 (2013)

\bibitem{pu2021automated}
Pu, J., Leader, J.K., Bandos, A., Ke, S., Wang, J., Shi, J., Du, P., Guo, Y.,
  Wenzel, S.E., Fuhrman, C.R., et~al.: Automated quantification of covid-19
  severity and progression using chest ct images. European Radiology
  \textbf{31}(1),  436--446 (2021)

\bibitem{roy2019quicknat}
Roy, A.G., Conjeti, S., Navab, N., Wachinger, C., Initiative, A.D.N., et~al.:
  Quicknat: A fully convolutional network for quick and accurate segmentation
  of neuroanatomy. NeuroImage  \textbf{186},  713--727 (2019)

\bibitem{shi2020radiological}
Shi, H., Han, X., Jiang, N., Cao, Y., Alwalid, O., Gu, J., Fan, Y., Zheng, C.:
  Radiological findings from 81 patients with covid-19 pneumonia in wuhan,
  china: a descriptive study. The Lancet infectious diseases  \textbf{20}(4),
  425--434 (2020)

\bibitem{wachinger2018deepnat}
Wachinger, C., Reuter, M., Klein, T.: Deepnat: Deep convolutional neural
  network for segmenting neuroanatomy. NeuroImage  \textbf{170},  434--445
  (2018)

\bibitem{wang2020noise}
Wang, G., Liu, X., Li, C., et~al.: A noise-robust framework for automatic
  segmentation of covid-19 pneumonia lesions from ct images. IEEE Transactions
  on Medical Imaging  \textbf{39}(8),  2653--2663 (2020)

\bibitem{wong2020frequency}
Wong, H.Y.F., Lam, H.Y.S., et~al.: Frequency and distribution of chest
  radiographic findings in patients positive for covid-19. Radiology
  \textbf{296}(2),  E72--E78 (2020)

\bibitem{wu_clinical_2020}
Wu, M.Y., Yao, L., Wang, Y., Zhu, X.Y., Wang, X.F., Tang, P.J., Chen, C.:
  Clinical evaluation of potential usefulness of serum lactate dehydrogenase
  ({LDH}) in 2019 novel coronavirus ({COVID}-19) pneumonia. Respir Res
  \textbf{21}(1), ~171 (Jul 2020)

\bibitem{zhang2019multiple}
Zhang, H., Valcarcel, A.M., Bakshi, R., Chu, R., Bagnato, F., Shinohara, R.T.,
  Hett, K., Oguz, I.: Multiple sclerosis lesion segmentation with tiramisu and
  2.5 d stacked slices. In: MICCAI. pp. 338--346 (2019)

\bibitem{zhang2020dabc}
Zhang, X., Yu, Z., Han, X., Zhao, B., Zhuo, Y., Ren, Y., Xue, X., Lamm, L.,
  Feng, J., Marr, C., et~al.: Dabc-net for robust pneumonia segmentation and
  prediction of covid-19 progression on chest ct scans  (2020)

\bibitem{zhou2020rapid}
Zhou, L., Li, Z., Zhou, J., et~al.: A rapid, accurate and machine-agnostic
  segmentation and quantification method for ct-based covid-19 diagnosis. IEEE
  transactions on medical imaging  \textbf{39}(8),  2638--2652 (2020)

\bibitem{zhou2020ct}
Zhou, S., Wang, Y., Zhu, T., Xia, L.: Ct features of coronavirus disease 2019
  (covid-19) pneumonia in 62 patients in wuhan, china. American Journal of
  Roentgenology  \textbf{214}(6),  1287--1294 (2020)

\bibitem{zhou_ct_2021}
Zhou, X., Pu, Y., Zhang, D., Xia, Y., Guan, Y., Liu, S., Fan, L.: {CT} findings
  and dynamic imaging changes of {COVID}-19 in 2908 patients: a systematic
  review and meta-analysis. Acta Radiol p. 284185121992655 (2021)

\end{thebibliography}

\end{document}